\documentclass{jetpl}
\usepackage{epsfig,bm,amsfonts,amssymb}
\twocolumn

\newcommand{\be}{\begin{equation}}
\newcommand{\ee}{\end{equation}}
\newcommand{\vex}{{\bm x}}
\newcommand{\vep}{{\bm p}}
\newcommand{\ver}{{\bm r}}
\newcommand{\veL}{{\bm L}}
\newcommand{\ven}{{\bm n}}
\newcommand{\vez}{{\bm z}}
\newcommand{\vesig}{{\bm\sigma}}
\newcommand{\verho}{{\bm \rho}}

\title{Gluonic correlation length from spin-dependent potentials}
\rtitle{Gluonic correlation length from spin-dependent potentials}

\author{A.\,M.\,Badalian, A.\,V.\,Nefediev, and Yu.\,A.\,Simonov}
\address{Institute of Theoretical and Experimental Physics,
117218, B.Cheremushkinskaya 25, Moscow, Russia}

\rauthor{A.\,M.\,Badalian, A.\,V.\,Nefediev, and Yu.\,A.\,Simonov}

\abstract{The vacuum gluonic correlation length is extracted from recent lattice data on spin-dependent interquark potentials in heavy quarkonia.
It is shown that the data are consistent with extremely small values of the correlation length, $T_g\lesssim 0.1$ fm.}


\begin{document}
\maketitle

\section{Introduction}

The history of the nonperturbative vacuum gluonic fields in QCD
can be started with the introduction of condensates in the
framework of the QCD sum rules \cite{1}, where the notion of the nonperturbative
vacuum fields was introduced and the first estimate of the
gluonic condensate, $G_2= \frac{\alpha_s}{\pi} \langle F^a_{\mu\nu}
F^a_{\mu\nu}\rangle=0.012$ GeV$^4$, was given. Higher condensates and
other possible vacuum averages of local operators, in line with
the Wilson expansion, were introduced and estimated in Ref.~\cite{2} (for a review see Ref.~\cite{3}).
However, for understanding the nonperturative dynamics of the QCD vacuum another characteristic is
vitally important --- namely, the vacuum correlation length of gluonic
fields $T_g$, which defines the nonlocality of gluonic excitations.
At a phenomenological level it was discussed in Ref.~\cite{4},
while its rigorous definition was given in the framework of the Field
Correlator Method (FCM)\footnote{This method is often called as the Stochastic Vacuum Model.} \cite{5,difdef} (see Ref.~\cite{6} for a review of
the method). The physical role of $T_g$ for the phenomenology of hadrons is quite important: in particular, for
hadrons of the spatial size $R$ and the temporal size $T_q$, the QCD sum rule method can be applied if $R,T_q\ll T_g$, while
potential-type approaches are valid in the opposite limit, $R,T_q\gg T_g$. An example of the exactly solvable theory with
$T_g\equiv 0$
is provided by the 't~Hooft model for QCD in two dimensions \cite{tHooft} (see Ref.~\cite{kn2} for a review), which reveals a lot of interesting
phenomena reminiscent of those one expects in four-dimensional QCD.

Direct lattice measurements of the gluonic correlation length give rather
small values for the latter, $T_g\approx 0.15\div 0.3$ fm \cite{Tg}. In this paper we adopt another strategy: using the FCM we extract the vacuum correlation
length from the spin-dependent potentials measured recently on the lattice, in the quenched approximation and without cooling \cite{Komas}.
We argue that the data prefer even smaller values of the vacuum correlation
length, $T_g\lesssim 0.1$ fm.
Notice that the spin-dependent potentials possess a direct physical meaning and are expected to be free of any artifacts of the definitions and methods
used. In particular, the lattice field strength correlators, also measured in Ref.~\cite{Komas}, are calculated as
matrix elements of operators defined in (potential) nonrelativistic QCD and simulated on the lattice. They are to be multiplied by
the appropriate renormalisation factor in order to give the corresponding correlators in continuum. Although such renormalisation factors are expected
to be calculated directly from QCD, they are, generally speaking, not known at present.

\section{Spin-dependent potentials in the Field Correlator Method}

Spin-dependent interaction in a heavy quarkonium is well known \cite{EF,EF2,Gr} and, to the order $O(1/m^2)$, it reads:
\begin{eqnarray}
V_{SD}(r)&=&\left(\frac{\vesig_1\veL}{4m_1^2r}+\frac{\vesig_2\veL}{4m_2^2r}\right)[V_0'(r)+2V_1'(r)]\nonumber\\
&+&\frac{(\vesig_1+\vesig_2)\veL}{2m_1m_2r}V_2'(r)+\frac{\vesig_1\vesig_2 }{12m_1m_2}V_4(r)\label{SO0}\\
&+&\frac{(3(\vesig_1\ven)(\vesig_2\ven)-\vesig_1\vesig_2)}{12m_1m_2}V_3(r)\nonumber,
\end{eqnarray}
where $m_i$ and $\vesig_i$ $(i=1,2)$ are the quark masses and spins, respectively. Primes denote the derivatives with respect to the interquark separation $r$.
The static interquark potential $V_0(r)$, together with the potentials $V_1(r)$ and $V_2(r)$, satisfies the Gromes relation \cite{Gr},
\be
V_0'(r)+V_1'(r)-V_2'(r)=0.
\label{Ge}
\ee
Notice that this relation refers both to the perturbative and nonperturbative parts of the potentials $V_n(r)$ $(n=0,1,2)$ and, while their perturbative parts
satisfy this relation identically, their nonperturbative parts satisfy it in the FCM. With the definition of $V_n(r)$ used in Ref.~\cite{EF} the
fulfilment of Eq.~(\ref{Ge}) is not evident. In all lattice calculations (see, for example, Refs.~\cite{Komas,18}) relation (\ref{Ge}) is satisfied only approximately.
For a detailed discussion of this issue see Ref.~\cite{S0}.

The spin-dependent interactions in the form of Eq.~(\ref{SO0}) can be derived naturally in the framework of the FCM \cite{11}. To this end
a quark--antiquark system is considered and its wave function is built:
$$
\Psi^{({\rm in, out})}_{q\bar q}(x,y|A)=\bar{\Psi}_{\bar q}(x)\Phi(x,y)\Psi_q(y),
$$
with nonlocal gauge invariance guaranteed by the parallel transporter,
$$
\Phi(x,y)=P\exp{\left(ig\int_{y}^{x}dz_{\mu}A_{\mu}^at^a\right)}.
$$
Then the Green's function can be constructed as
\begin{eqnarray*}
G_{q\bar q}&=&\langle\Psi_{q\bar q}^{({\rm out})}(\bar{x},\bar{y}|A)
\Psi^{({\rm in})\dagger}_{q\bar q}(x,y|A)\rangle_{q\bar{q}A}\\[2mm]
&=&\langle {\rm Tr}S_q(\bar{x},x|A)\Phi(x,y)S_{\bar{q}}(y,\bar{y}|A)\Phi(\bar{y},\bar{x})\rangle_A,
\end{eqnarray*}
where $S_{q}$ and $S_{\bar{q}}$ are the propagators of the quark and the antiquark, respectively, in the background gluonic field.
Using the Feynman--Schwinger representation for the single-quark propagators one can see that the interquark interaction is described
in terms of the Wilson loop $W(C)$, with the contour $C$ running over the quark trajectories, averaged over the background gluonic field.
More specifically, the value $\langle{\rm Tr}W(C)\rangle$, which enters the quarkonium Greens' function, can be expressed through the correlators of the
field strength tensors as
\begin{eqnarray*}
\langle{\rm Tr}W(C)\rangle&=&\left\langle{\rm Tr}\exp ig\int d\pi_{\mu\nu}(z)F_{\mu\nu}(z)\right\rangle\\
=\exp\sum^\infty_{n=1}\frac{(ig)^n}{n!}&\displaystyle\int& d\pi(1)\ldots\int d \pi(n)\langle\langle F(1)\ldots F(n)\rangle\rangle,
\end{eqnarray*}
where the cluster expansion theorem \cite{S0,18} was used. The average $\langle\langle\ldots\rangle\rangle$
stands for connected correlators, for example, for the bilocal correlator,
$\langle\langle F(1)F(2)\rangle\rangle=\langle F(1)F(2)\rangle-\langle F(1)\rangle\langle F(2)\rangle$,
and $F_{\mu\nu}=\partial_\mu A_\nu-\partial_\nu A_\mu- ig [A_\mu, A_\nu]$
is the vacuum field strength. Obviously, $\langle\langle F\rangle\rangle=\langle F\rangle=0$.

The element of integration,
$$
d\pi_{\mu\nu}(z)=ds_{\mu\nu}(z)-i\sigma^{(1)}_{\mu\nu}d\tau_1+i\sigma_{\mu\nu}^{(2)}d\tau_2,
$$
with $\sigma_{\mu\nu}^{(i)}=\frac{1}{4i}(\gamma_\mu\gamma_\nu-\gamma_\nu\gamma_\mu)$ ($i=1,2$ for the quark and antiquark, respectively),
contains both the element of the surface area and the
quark spin variables accompanied by the quark proper time differentials.
This is the most economical way to include spin-dependent interactions into consideration \cite{Lisbon}.
In the Gaussian approximation for the vacuum, when only the lowest, bilocal correlator
is retained (see Ref.~\cite{30} for the discussion of the accuracy of this approximation), one has
\be
\langle Tr W\rangle\propto\exp\left[-\frac12\int_Sd\pi_{\mu\nu}(x)d\pi_{\lambda\rho}(x')
D_{\mu\nu\lambda\rho}(x-x')\right],
\label{W2}
\ee
where
$$
D_{\mu\nu\lambda\rho}(x-x')\equiv\frac{g^2}{N_c}\left\langle{\rm Tr}F_{\mu\nu}(x)\Phi(x,x')F_{\lambda\rho}(x')\Phi(x',x)\right\rangle.
$$
This bilocal correlator can be expressed through only two gauge-invariant scalar functions, $D(z)$ and $D_1(z)$ \cite{5,difdef}:
\begin{eqnarray*}
D_{\mu\nu\lambda\rho}(z)=(\delta_{\mu\lambda}\delta_{\nu\rho}&-&\delta_{\mu\rho}\delta_{\nu\lambda})D(z)\\[3mm]
+\frac12\left[\frac{\partial}{\partial z_\mu}(z_\lambda\delta_{\nu\rho}\right.&-&\left.z_\rho\delta_{\lambda\nu})+
\genfrac{(}{)}{0pt}{0}{\mu\leftrightarrow\nu}{\lambda\leftrightarrow\rho}\right]D_1(z).\nonumber
\end{eqnarray*}

The correlator $D(z)=D(z_0,|\vez|)$ contains only a nonperturbative part and it is responsible for the QCD string formation at large interquark
separations. The fundamental string tension can be expressed as a double integral:
\be
\sigma=2\int_0^\infty d\nu\int_0^\infty d\lambda D(\nu,\lambda).
\label{sigma}
\ee
This correlator decreases in all directions of the Euclidean space, and this decrease is governed by the gluonic correlation
length $T_g$,
\be
D(z)=\frac{\sigma}{\pi T_g^2}\exp[-z/T_g],\quad z=\sqrt{\lambda^2+\nu^2},
\label{Dcor1}
\ee
where the coefficient is chosen to satisfy the relation (\ref{sigma}). Strictly speaking, the asymptotic
form of the correlator (\ref{Dcor1}) is valid at large distances, at
$|z|\gtrsim T_g$, while at $|z|\lesssim T_g$ the function $D(z)$ admits a Taylor expansion with the parameter $z^2/T_g^2$ (see, for
example, Ref.~\cite{BY}).
Notice however that, for small $T_g$, the contribution of the region $|z|\lesssim T_g$ to the integrals defining the spin-dependent potentials
(see Eq.~(\ref{Vseq}) below) is negligible, so we stick to the simplified form (\ref{Dcor1}) from the beginning.

Another important comment concerning the correlation length $T_g$ is the scale at which it is defined (see, for example, Ref.~\cite{BM1} for the
discussion of the issue in relation to the Operator Product Expansion in QCD).
By natural arguments we expect the scale of $T_g$ to be
of order of the average size of gluelump (average momentum), which is of
the order of 1 GeV (see Ref.~\cite{gluelump}).

The other correlator, $D_1(z)$, contains both perturbative and nonperturbative contributions. Its perturbative part leads to the colour Coulomb
interaction between the quarks, whereas its nonpeturbative part was parametrised through the gluelump spectrum in Ref.~\cite{simcor}. We use this
parametrisation here:
$$
D_1(z)=\frac{4\alpha_S C_F}{\pi}\frac{1}{z^4}+\frac{2 C_F\alpha_S\sigma_{\rm adj}}{T_g' z}\exp[-z/T_g'],
$$
where $C_F=(N_C^2-1)/(2N_C)=4/3$, $C_{\rm adj}=N_C=3$, $\sigma_{\rm adj}=(C_{\rm adj}/C_F)\sigma=(9/4)\sigma$, and $\alpha_S=g^2/(4\pi)$. Then
$$
D_1(z)=\frac{16\alpha_S}{3\pi}\frac{1}{z^4}+\frac{6\alpha_S\sigma}{zT_g'}\exp[-z/T_g'].
$$
For the sake of generality the correlation length $T_g'$ is kept different from $T_g$ --- see Ref.~\cite{simcor} for the details.

The spin-independent part of the interquark interaction follows from the area law asymptotic for the Wilson loop (\ref{W2}),
$$
\langle Tr W(C)\rangle\sim \exp{(-\sigma S_{\rm min}(C))},
$$
where $S_{\rm min}$ is the area of the minimal surface defined by the contour $C$,
\be
S_{\rm min}(C)=\int_0^Tdt\int_0^1d\beta\sqrt{(\dot{w}w')^2-\dot{w}^2w'^2},
\label{Smin}
\ee
with the string profile function $w_\mu(t,\beta)$ approximated by the
straight--line ansatz:
\be
w_{\mu}(t,\beta)=\beta x_{1\mu}(t)+(1-\beta)x_{2\mu}(t).
\label{anz}
\ee
Here $x_{1,2}(t)$ are the four--coordinates of the quarks at the ends of the string.
This approximation is valid at least for not large excitations due to
the fact that hybrid excitations responsible for the string
deformation are decoupled from mesonic excitations by the mass gap of order 1~GeV.

The spin-dependent interquark interaction follows
from the mixed terms of the form $ds_{\mu\nu}\sigma_{\lambda\sigma}F_{\mu\nu}F_{\lambda\sigma}d\tau$ in Eq.~(\ref{W2}),
\begin{eqnarray*}
L_{SO}=\int d s_{\mu\nu}(w)d\tau_1\sigma^{(1)}_{\lambda\rho}D_{\mu\nu\lambda\rho}(w-x_1)+(1\to 2),\\
ds_{\mu\nu}=\varepsilon^{ab}\partial_a w_\mu(t,\beta)\partial_b w_\nu(t,\beta)dtd\beta,\quad a,b=\{t,\beta\}.
\end{eqnarray*}
For the ansatz (\ref{anz}) one has:
\begin{eqnarray*}
ds_{i4}&=&r_idtd\beta,\quad ds_{ik}=\varepsilon_{ikm}\rho_mdtd\beta,\\[2mm]
\verho&=&[\ver\times(\beta\dot{\vex}_1+(1-\beta)\dot{\vex}_2)],
\end{eqnarray*}
and thus the angular momentum enters the interaction through $ds_{ik}$.
Finally, we introduce the laboratory time $t$ instead of the proper quark times as
$d\tau_i=dt/(2\mu_i)$ ($i=1,2$), where $\mu_i$ stands for the $i$-th quark energy \cite{DKS,12}.
Notice that the proper inertia of the string is to be taken into account
when proceeding from the quark velocities to the angular momentum variables \cite{BNS1}.
However, the effect of the string inertia and the
deviation of the quark energy from its mass are important for light quarks, whereas they are negligible in case of heavy quarks, so that,
for the heavy quarkonium, one has simply
$\mu_i=m_i$ and $\vep_i=m_i\dot{\vex}_i$. Then, by an explicit calculation,
one can arrive at Eq.~(\ref{SO0}) (the details of the derivation can be found in Refs.~\cite{Lisbon,242}) with the following identification
of the potentials $V_n(r)$, $n=0$-4:
\begin{eqnarray}
V_0'(r)&=&2\int_0^\infty d\nu\int_0^rd\lambda D(\lambda,\nu)+r\int_0^\infty d\nu D_1(r,\nu),\nonumber\\
V_1'(r)&=&-2\int_0^\infty d\nu\int_0^r d\lambda \left(1-\frac{\lambda}{r}\right)D(\lambda,\nu),\label{Vseq}\\
V_2'(r)&=&\frac{2}{r}\int_0^\infty d\nu\int_0^r \lambda d\lambda D(\lambda,\nu)+r\int_0^\infty d\nu D_1(r,\nu),\nonumber\\
V_3(r)&=&-2r^2\frac{\partial}{\partial r^2}\int_0^\infty d\nu D_1(r,\nu),\nonumber\\
V_4(r)&=&6\int_0^\infty d\nu\left[D(r,\nu)+\left[1+\frac23r^2\frac{\partial}{\partial\nu^2}\right]D_1(r,\nu)\right].\nonumber
\end{eqnarray}
At large interquark separations, $r\gg T_g,T_g'$, these yield simply:
\begin{eqnarray}
V_0'(r)=\sigma+\frac43\frac{\alpha_S}{r^2},\quad V_1'(r)&=&-\sigma,\quad V_2'(r)=\frac43\frac{\alpha_S}{r^2},\nonumber\\[-1mm]
\label{V012eq}\\[-1mm]
V_3(r)=\frac{4\alpha_S}{r^3},\quad V_4(r)&=&\frac{32}{3}\pi\alpha_S\delta^{(3)}(r)\nonumber.
\end{eqnarray}
In particular, the static interquark potential comes out from Eq.~(\ref{V012eq}) in the standard ``linear+Coulomb" form,
$$
V_0(r)=V_{Q\bar{Q}}(r)=\sigma r-\frac43\frac{\alpha_S}{r}+{\rm const}.
$$

\section{Results and discussion}

\begin{figure}[ht]
\begin{center}
\epsfig{file=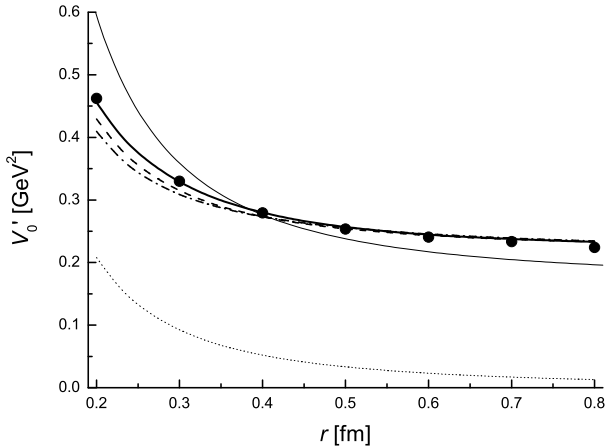,width=8cm}
\caption{The profile of the $V_0'(r)$ for the fit~1 (dash-dotted line), fit~2 (dashed line), fit~3 (fat solid line), fit~4 (doted line),
and fit~5 (thin solid line). Lattice data are given by dots.}\label{V0fig}
\end{center}
\end{figure}

\begin{figure}[ht]
\begin{center}
\epsfig{file=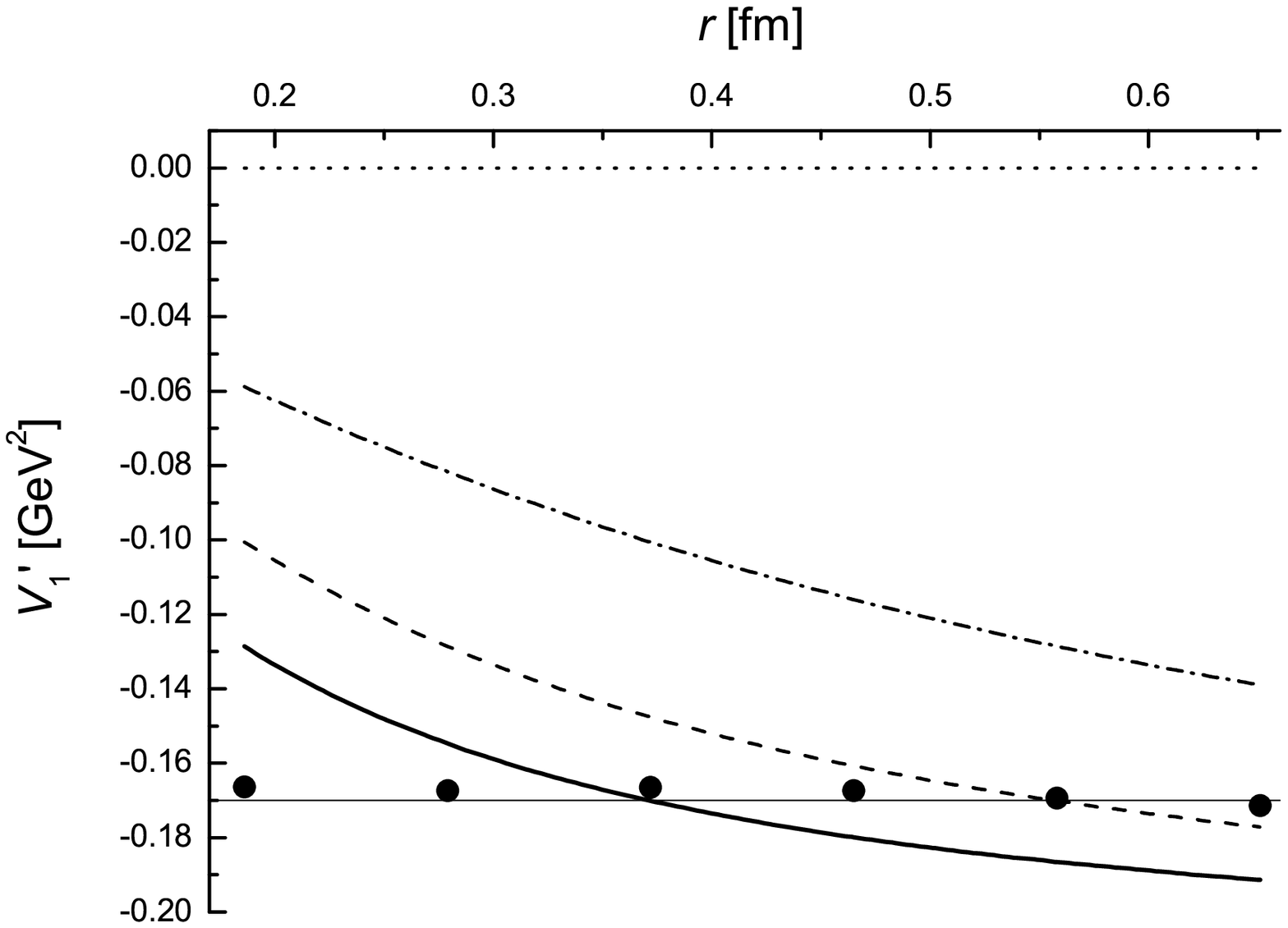,width=8cm}
\caption{The same as in Fig.~\ref{V0fig} but for $V_1'(r)$.}\label{V1fig}
\end{center}
\end{figure}

\begin{figure}[ht]
\begin{center}
\epsfig{file=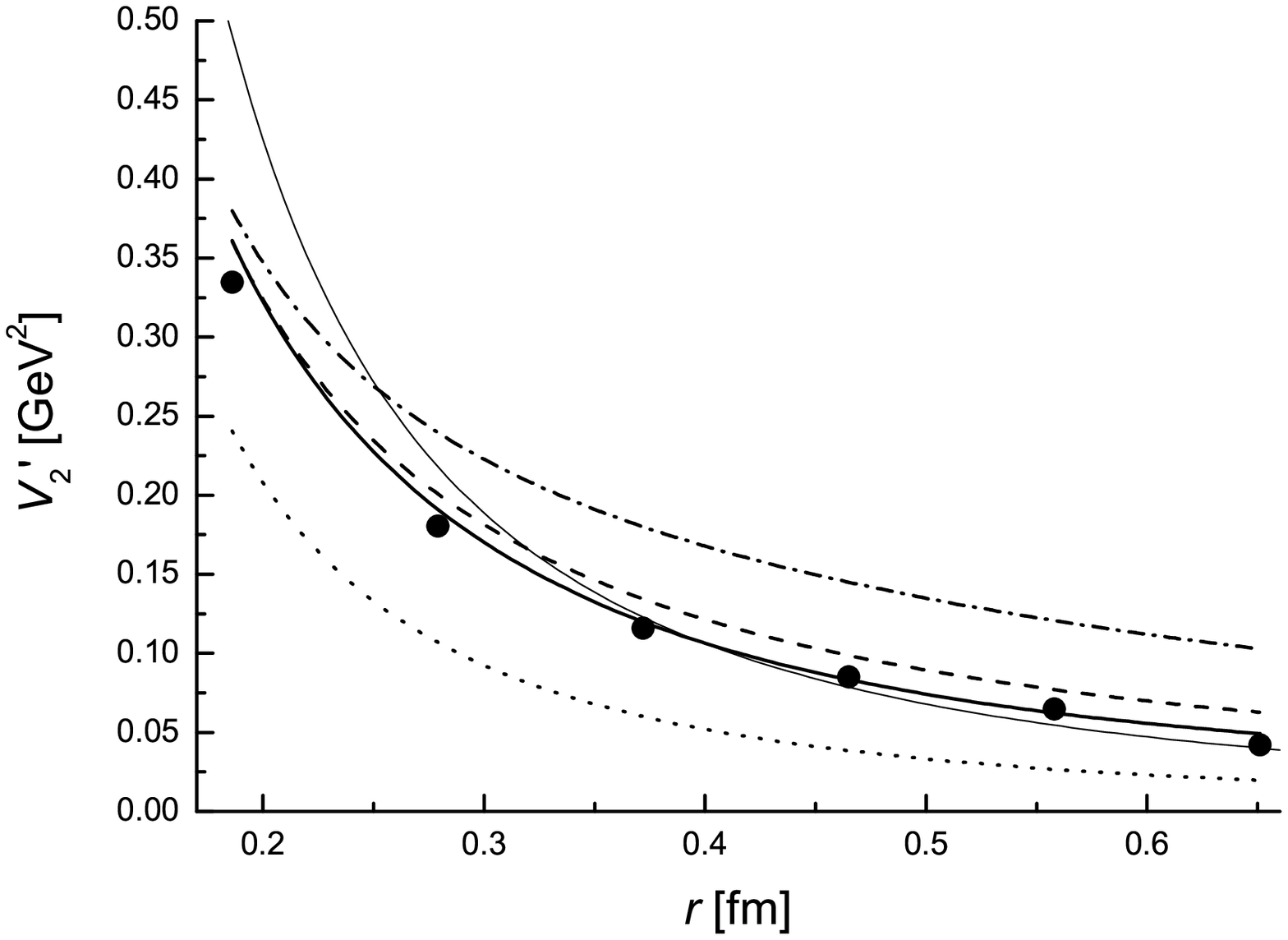,width=8cm}
\caption{The same as in Fig.~\ref{V0fig} but for $V_2'(r)$.}\label{V2fig}
\end{center}
\end{figure}

\begin{figure}[ht]
\begin{center}
\epsfig{file=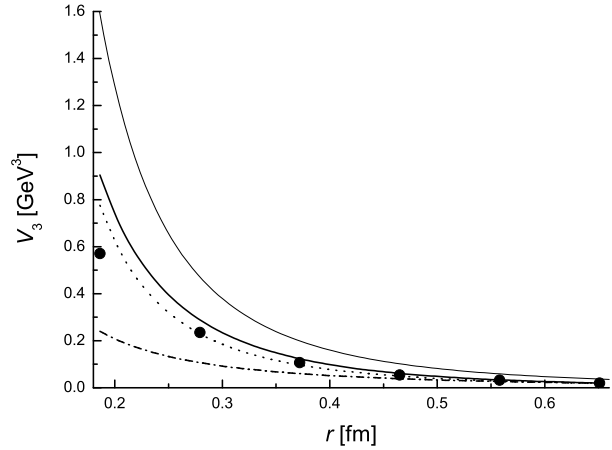,width=8cm}
\caption{The same as in Fig.~\ref{V0fig} but for $V_3(r)$. The curve for the fit~2 coincides with that for the fit~1.}\label{V3fig}
\end{center}
\end{figure}

\begin{figure}[ht]
\begin{center}
\epsfig{file=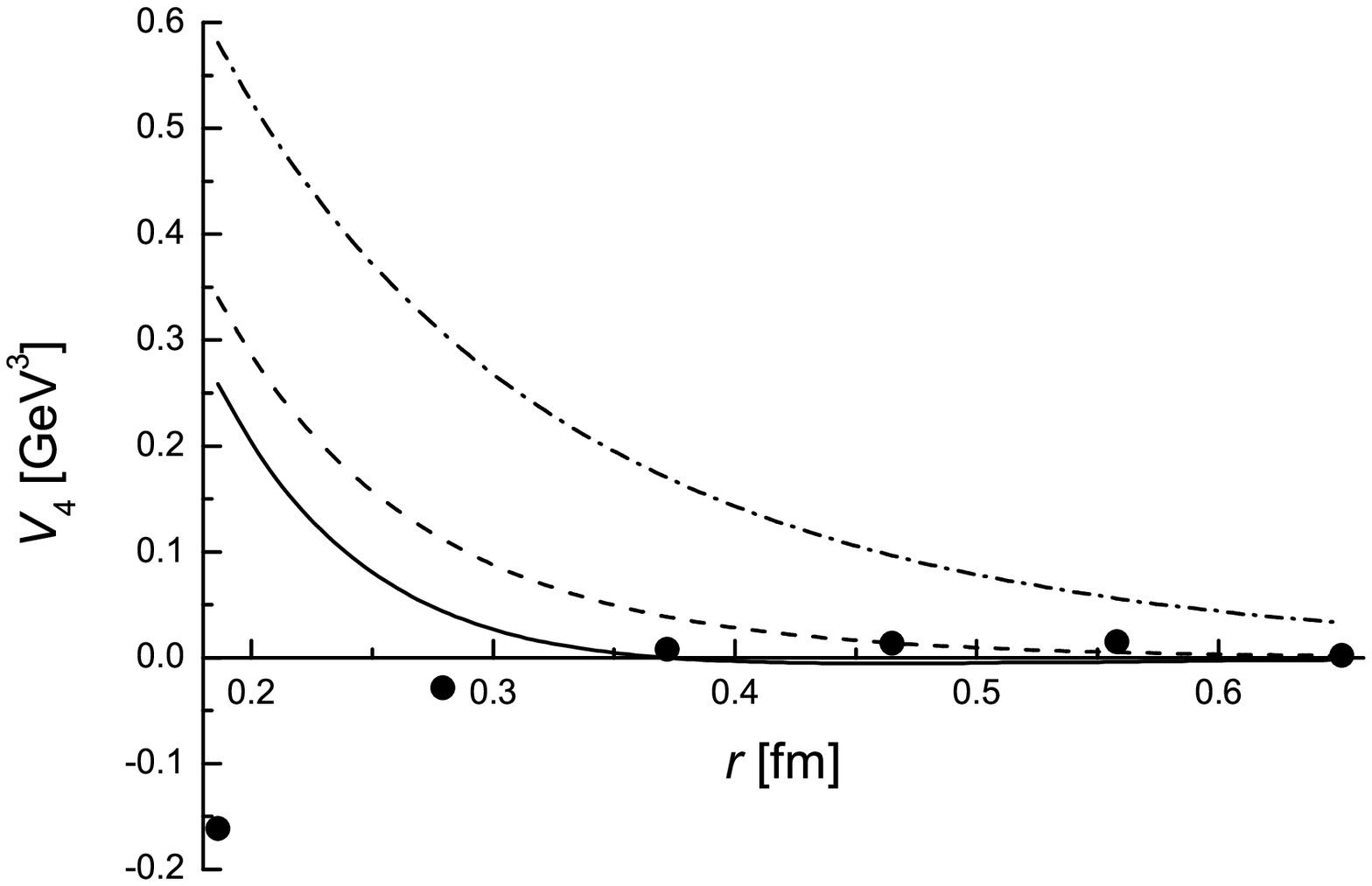,width=8cm}
\caption{The same as in Fig.~\ref{V0fig} but for $V_4(r)$. The fits~4 and 5 correspond to
delta-functions localised at $r=0$.}\label{V4fig}
\end{center}
\end{figure}

With the form of the potentials (\ref{Vseq}) we are in a position to fit the corresponding lattice data from Ref.~\cite{Komas}.
We have the set of four fitting parameters: $\{\alpha_S,\sigma,T_g,T_g'\}$. In Table~1 we give the set of our fits. For the fits~1-3 we use
the full form of the potentials (\ref{Vseq}), the fit~4 demonstrates the relevance of nonperturbative interactions since only
perturbative part of the potentials $V_n(r)$ ($n=0$-4) is retained in this case. Finally, the data were fitted with the help of the asymptotic
large--distance potentials (\ref{V012eq}) (fit~5). Comparison of our fits with the lattice data is given by Figs.~\ref{V0fig}-\ref{V4fig}
(we use the data from Ref.~\cite{Komas} for the $20^3 40$ lattice with the coupling $\beta=6.0$ and the lattice spacing $a=0.093$ fm).

One can draw several conclusions from Figs.~\ref{V0fig}-\ref{V4fig}: i) the data clearly indicates the presence of nonperturbative
contributions to the potentials even at small interquark separations, where the perturbative physics dominates;
ii) although the simplified fit~5 approximates the data rather well, the fits~1-3 give a better description of the latter. Thus one concludes that
the present data allow one to study the ``anatomy" of the field correlators;
iii) by comparing fits~1-3 with one another one can conclude that the data prefer small correlation lengths $T_g$ and $T_g'$, a good description
is achieved with them both being 0.1~fm or below;
iv) the theoretical expressions for $V_n(r)$ $(n=0,1,2)$ given by our Eqs.~(\ref{Vseq}) satisfy the Gromes relation (\ref{Ge}) identically,
while the lattice potentials, as was stressed in Ref.~\cite{Komas}, violate it. The source of this violation is not clear at the moment: it may be
a lattice artifact and disappear in the continuum limit \cite{Komapriv} or it may be related to the contribution of higher-order correlators
taken into account
differently in different potentials entering the Gromes relation \cite{difdef}. This question deserves an additional careful investigation. In particular, a detailed
comparison of the Eichten--Feinberg and Field Correlator definitions of these potentials is given in Ref.~\cite{BNSprep};
v) there is a certain contradiction between the theoretical predictions and the lattice data for the potential $V_4(r)$. Our form of
the potential $V_4(r)$ given by Eq.~(\ref{Vseq}) is consistent with the delta-functional form of this potential in the limit of the vanishing
correlation length (see Eq.~(\ref{V012eq})). Although, for finite values of the correlation length, $V_4(r)$ is smeared and can become negative,
the small-$r$ behaviour of our
fits is essentially different from that of the lattice data, so this question deserves additional investigation.

Notice that the values of the correlations lengths obtained in the fits to the data are in good agreement with the predictions made for the
gluelump spectrum in the framework of the FCM \cite{simcor}. Indeed, the inverse of the correlation length gives the mass of the lowest gluelump
in the corresponding channel. From the best fit~3 these masses constitute 3 GeV and 2 GeV for the correlators $D(z)$ and $D_1(z)$, respectively.
These should be confronted with the predictions 2.8 GeV and 1.7 GeV made in Ref.~\cite{simcor} (after the proper rescaling from
$\sigma=0.18$ GeV$^2$ used in Ref.~\cite{simcor} to $\sigma=0.22$ GeV$^2$ extracted from out fits).

The fact that the vacuum correlation length is small, less than 0.1~fm, allows one to justify the use of the so-called string limit of QCD, when this length
is set equal to zero. In this limit, the interaction of colour constituents in hadrons can be described in terms of the infinitely thin QCD string with
the string action given by Eq.~(\ref{Smin}). This approach was
successfully used to describe conventional mesons \cite{12,mesons} and baryons \cite{baryons}, hybrids \cite{hybrid}, glueballs \cite{glbl}, and
gluelumps \cite{gluelump}.

We conclude that the existing lattice data on the spin-dependent potentials in heavy quarkonia are consistent with the predictions of the FCM and that the gluonic correlation
length extracted from these data is small,
less that 0.1 fm. The definition and the behaviour of the lattice potentials $V_1(r)$ and $V_4(r)$ might need a better justification and, possibly, some
improvements.
\smallskip

The authors are grateful to N. Brambilla, A. Vairo, and Y. Koma for useful discussions. This work was supported by the Federal Agen\-cy for Atomic Energy of Russian Fe\-de\-ration, by the Federal Programme of
the Russian Ministry of Industry, Science, and Technology No. 40.052.1.1.1112, and by the grant NSh-4961.2008.2 for the leading scientific
schools. A.B. and Yu.S. would like to acknowledge the financial support through the grants RFFI-06-02-17012 and RFFI-06-02-17120.
Work of A.N. was supported by RFFI-05-02-04012-NNIOa, DFG-436 RUS 113/820/0-1(R), PTDC/FIS/70843/2006-Fisica,
and by the non-profit ``Dynasty" foundation and ICFPM.

\begin{table}[t]
\begin{center}
\begin{tabular}{|c|c|c|c|c|}
\hline
&$\alpha_S$&$\sigma$, GeV$^2$&$T_g$, fm&$T_g'$, fm\\
\hline
fit 1&0.16&0.22&0.2&0.2\\
\hline
fit 2&0.16&0.22&0.1&0.1\\
\hline
fit 3&0.16&0.22&0.07&0.1\\
\hline
fit 4&0.16&$-$&0&0\\
\hline
fit 5&0.32&0.17&$-$&$-$\\
\hline
\end{tabular}
\caption{The set of fits for the lattice data on spin-dependent potentials taken from Ref.~\cite{Komas}.
Eqs.~(\ref{Vseq}) and (\ref{V012eq}) are used for the fits~1-4 and fit~5, respectively.}
\end{center}
\end{table}

\begin{thebibliography}{99}
\bibitem{1} M. A. Shifman, A. I. Vainshtein, and V. I. Zakharov, Nucl. Phys. B {\bf 147}, 385 , 448 (1979); Nucl. Phys. B {\bf 147}, 448 (1979).
\bibitem{2} V. A. Novikov, M. A. Shifman, A. I. Vainshtein, and V. I. Zakharov, Phys. Lett. B {\bf 86}, 347 (1979).
\bibitem{3} B. L. Ioffe, Prog. Part. and Nucl. Phys. {\bf 56}, 232 (2006).
\bibitem{4} M. B. Voloshin, Nucl. Phys. B {\bf 154}, 365  (1979); Yad. Fiz {\bf 36}, 247 (1982) (Sov. J. Nucl. Phys. {\bf 36}, 143 (1982));
D. Gromes, Phys. Lett. B {\bf 115}, 482 (1982); V. Marquard and H. G. Dosch, Phys. Rev. D {\bf 35}, 2238 (1987).
\bibitem{5} H. G. Dosch, Phys. Lett.  B {\bf 190}, 177 (1987); H. G. Dosch and Yu. A. Simonov, Phys. Lett.  B {\bf 205}, 339 (1988).
\bibitem{difdef} Yu. A. Simonov, Nucl. Phys. B {\bf 307}, 512 (1988).
\bibitem{6} A. Di Giacomo, H. G. Dosch, V. I. Shevchenko, and Yu. A. Simonov, Phys. Rep. {\bf 372}, 319 (2002).
\bibitem{tHooft} G. 't Hooft, Nucl. Phys. B {\bf 72}, 461 (1974); B {\bf 75}, 461 (1974); I. Bars and M. B. Green, Phys. Rev. D {\bf 17}, 537 (1978).
\bibitem{kn2} Yu. S. Kalashnikova and A. V. Nefediev, Usp. Fiz. Nauk {\bf 172}, 378 (2002) (Phys. Usp. {\bf 45}, 347 (2002)).
\bibitem{Tg} M. Campostrini, A. Di Giacomo, and G. Mussardo, Z. Phys. C {\bf 25}, 173 (1984);
M. Campostrini, A. Di Giacomo, and S. Olejnik, Z. Phys. C {\bf 34}, 577 (1986);
G. Bali, N. Brambilla, A. Vairo, Phys. Lett. B {\bf 421}, 265 (1998).
\bibitem{Komas} Y. Koma and M. Koma, Nucl. Phys. B {\bf 769}, 79 (2007).
\bibitem{EF} E. Eichten and F. L. Feinberg, Phys. Rev. D {\bf 23}, 2724 (1981).
\bibitem{EF2} A. Barchielli, N. Brambilla, and G. M. Prosperi, Nuovo Cim. A {\bf 103}, 59 (1990);
A. Pineda and A. Vairo, Phys. Rev. D {\bf 63}, 054007 (2001); N. Brambilla and A. Vairo, Phys. Rev. D {\bf 55}, 3974 (1997).
\bibitem{Gr} D. Gromes, Z. Phys. C {\bf 26}, 401 (1984).
\bibitem{18} K. D. Born, E. Laermann, T. F. Walsh, and P. M. Zerwas, Phys. Lett. B {\bf 329}, 332 (1994);
G. S. Bali, K. Schilling, and A. Wachter, Phys. Rev. D {\bf 56}, 2566 (1997); Phys. Rev. D {\bf 55}, 5309 (1997).
\bibitem{S0} Yu. A. Simonov, Nucl. Phys. B {\bf 324}, 67 (1989).
\bibitem{11} Yu. A. Simonov, Yad. Fiz. {\bf 58} (1995) 113 (Phys. At. Nucl. {\bf 58}, 309 (1995)).
\bibitem{Lisbon} Yu. A. Simonov, in {\it Proceedings of the XVII International School
of Physics \lq\lq QCD: Perturbative or Nonperturbative,"} Lisbon, 1999, ed. by
L. S. Ferreira, P. Nogueira, and J. I. Silva-Marcos (World Scientific 2000), p. 60.
\bibitem{30} V. I. Shevchenko and Yu. A. Simonov, Phys. Rev. Lett. {\bf 85}, 1811 (2000);
Int. J. Mod. Phys. A {\bf 18}, 127 (2003); G. S. Bali, Phys. Rev. D {\bf 62}, 114503 (2000);
S. Deldar, Phys. Rev. D {\bf 62}, 034509 (2000).
\bibitem{BY}  A. M. Badalian and V. P. Yurov, Yad. Fiz. {\bf 51}, 1869 (1990).
\bibitem{BM1} A. P. Bakulev and S. V. Mikhailov, Phys. Rev. D {\bf 65}, 114511 (2002).
\bibitem{gluelump} Yu. A. Simonov, Nucl. Phys. B {\bf 592}, 350 (2001).
\bibitem{simcor} Yu. A. Simonov, Phys. At. Nucl. {\bf 69}, 528 (2006).
\bibitem{DKS} A. Yu. Dubin, A. B. Kaidalov, and Yu. A. Simonov, Phys. Lett. B {\bf 323}, 41 (1994).
\bibitem{12} Yu. S. Kalashnikova,  A. V. Nefediev, and Yu. A. Simonov, Phys. Rev. D {\bf 64}, 014037 (2001).
\bibitem{BNS1} A. M. Badalian, A. V. Nefediev, and Yu. A. Simonov, Pis'ma v ZhETF, {\bf 88}, 179 (2008).
\bibitem{242} Yu. A. Simonov, in {\em "Sense of Beauty in Physics"}, volume in honour of Adriano Di Giacomo, Pisa
Univ. Press, 2006, p.29; arXiv:hep-ph/0512242; A. M. Badalian and Yu. A. Simonov, Yad. Fiz. {\bf 59}, 2247 (1996)
(Phys. At. Nucl. {\bf 59}, 2164 (1996)).
\bibitem{Komapriv} Y. Koma, private communication.
\bibitem{BNSprep} A. M. Badalian, A. V. Nefediev, and Yu. A. Simonov, in preparation.
\bibitem{mesons} A. M. Badalian and B. L. G. Bakker, Phys. Rev. D {\bf 66}, 034025 (2002), Phys. Rev. D {\bf 70}, 016007 (2004).
\bibitem{baryons} Yu. A. Simonov, Phys. Rev. D {\bf 65}, 116004 (2002); I. M. Narodetskii and M. A. Trusov, Yad. Fiz. {\bf 65}, 949 (2002)
(Phys. At. Nucl. {\bf 65}, 917 (2002));
Yad. Fiz. {\bf 67}, 783 (2004) (Phys. At. Nucl. {\bf 67}, 762 (2004)); I. M. Narodetskii, C. Semay, and A. I. Veselov, arXive:0801.4270.
\bibitem{hybrid} Yu. A. Simonov, in {\it Proceedings of Hadron'93}, Como, Ed. by T. Bressani, A. Felicielo, G. Preparata, and P. G. Ratcliffe (Nuovo
Cim. A {\bf 107}, 2629 (1994)); Yu. S. Kalashnikova and Yu. B. Yufryakov,
Phys. Lett. B {\bf 359}, 175 (1995); Yad. Fiz. {\bf 60}, 374 (1997)
(Phys. At. Nucl. {\bf 60}, 307 (1997)); Yu. S. Kalashnikova and A. V. Nefediev, Phys. Rev. D {\bf 77}, 054025 (2008).
\bibitem{glbl} Yu. A. Simonov and A. B. Kaidalov, Phys. Lett. B {\bf 477}, 163
(2000); Yad. Fiz. {\bf 63}, 1507 (2000) (Phys. At. Nucl. {\bf 63}, 1428 (2000)).

\end{thebibliography}
\end{document}